\documentclass[aps,twocolumn,prd,showpacs,nofootinbib]{revtex4}
\usepackage{amsmath}
\usepackage{graphicx}
\usepackage{dcolumn}
\usepackage{bm}
\usepackage{amssymb}
\usepackage{latexsym}

\def\be{\begin{equation}}
\def\ee{\end{equation}}
\def\ba{\begin{eqnarray}}
\def\ea{\end{eqnarray}}

\begin{document}

\title{ Gravitational Wave Background from Phantom Superinflation
}

\author{Yun-Song Piao$^{a,b}$}
\affiliation{${}^a$College of Physical Sciences, Graduate School
of Chinese Academy of Sciences, YuQuan Road 19{\rm A}, Beijing
100049, China} \affiliation{${}^b$Interdisciplinary Center of
Theoretical Studies, Chinese Academy of Sciences, P.O. Box 2735,
Beijing 100080, China}

\begin{abstract}
Recently, the early superinflation driven by phantom field has
been proposed and studied. The detection of primordial
gravitational wave is an important means to know the state of very
early universe. In this brief report we discuss in detail the
gravitational wave background excited during the phantom
superinflation.
\end{abstract}

 \maketitle

Now a number of detectors for gravitational waves are (or expected
to start) operating, one of whose aim is searching a stochastic
background of gravitational wave. This background is expected to
have different components with different origins. The primordial
gravitational wave among them will be especially interesting,
since it would carry information about the state of the very early
universe. The basal mechanism of generation of primordial
gravitational wave in cosmology has been discussed in Ref.
\cite{G, A}, see \cite{S, RSV, FP, S1, AW, AH} for the stochastic
gravitational wave generated during the inflation, \cite{GG, BGGV,
BMU} for that of Pre Big Bang scenario, \cite{BST} for that of
cyclic model.

Recently, the superinflation driven by the phantom field, in which
the parameter of state equation $w< -1$ and the null energy
condition is violated \footnote{ Note that it is also possible to
get phantom energy in scalar-tensor theories of gravity without
non canonical kinetic terms \cite{BEPS}.} , has been proposed and
studied in Ref. \cite{PZ, PZ1, GJ, NO, ABV, BFM}.
The initial perturbation during the phantom superinflation will
leave the horizon, and can reenter the horizon during the
radiation/matter domination, which may be regarded as our late
time observable universe.
The phantom superinflation model leads to a blue or strong blue
spectrum of primordial tensor perturbations, which is different
from a nearly scale invariant (slightly red) spectrum predicted by
the inflation model. The reason is that the phantom superinflation
involves an early phase of super acceleration in which the Hubble
parameter increases with the time, while the inflation does not.
Thus it may be essential to revisit in detail the gravitational
wave background from the phantom superinflation.

We set $m_p^2 =1$ and work with the parameter $\epsilon
\equiv-{\dot h}/h^2$ being constant for simplicity, where $h\equiv
{\dot a}/a$. We firstly briefly show the model-independent
characters of phantom superinflation. The evolution of scale
factor during the phantom phase can be simply taken as $ a(t) \sim
(-t)^{n}$ \cite{PZ2}, where $t$ is from $-\infty$ to $0_-$, and
$n$ is a negative constant. Thus we have $\epsilon= 1/n<0$. In the
conformal time,
we obtain $-\eta\sim (- t)^{-n+1}$, and thus \be a(\eta) \sim
(-\eta)^{{n\over 1-n}}\equiv (-\eta)^{1\over \epsilon -1};
\label{aeta}\ee \be h \equiv {a^\prime \over a^2}= {1\over
(\epsilon -1) a \eta}, \label{htau}\ee where the prime denotes the
derivative with respect to $\eta$. The perturbations leaving the
horizon during the phantom superinflation can reenter the horizon
during radiation/matter domination, which may be responsible for
the structure formation of our observable universe. Further it may
be convenient to define \be {\cal N} \equiv \ln({a_e h_e \over a_i
h_i}), \label{caln}\ee which measures the efolding number that the
perturbation with the present horizon scale leaves the horizon
before the end of the phantom phase, where the subscript e and i
denote the end time of the phantom superinflation and the time
that the perturbation with the present horizon scale leaves the
horizon, respectively, and thus $a_ih_i=a_0h_0$, where the
subscript 0 denotes the present time, see Ref. \cite{KST}. From
(\ref{aeta}) and (\ref{htau}), we obtain \be a\sim \left({1\over
(1-\epsilon) a h}\right)^{1\over \epsilon -1}.\label{ah}\ee Thus
we have \be {a_e\over a_i}=({a_i h_i\over a_e h_e})^{1\over
\epsilon -1}= e^{{{\cal N}\over 1-\epsilon}} . \label{asim}\ee We
can see that during the phantom superinflation, the change of
scale factor is dependent on $\epsilon$. For the negative enough
$\epsilon$, the change $\Delta a/a=(a_e-a_i)/a_i\simeq {\cal
N}/(1-\epsilon)$ of $a$ can be very small. Taking the logarithm in
both sides of (\ref{ah}), we obtain \footnote{The equation can be
shown and actually also applied for the expansion with arbitrary
constant $\epsilon$. } \be \ln{({1\over ah})}=(\epsilon -1)\ln{a}
. \label{ahi} \ee We plot Fig.1 to further illustrate the
characters of the phantom superinflation. We assume, throughout
this brief report, that after the phantom phase ends, the
reheating will rapidly occur and then bring the universe back to
the usual FRW evolution.

\begin{figure}[t]
\begin{center}
\includegraphics[width=7cm]{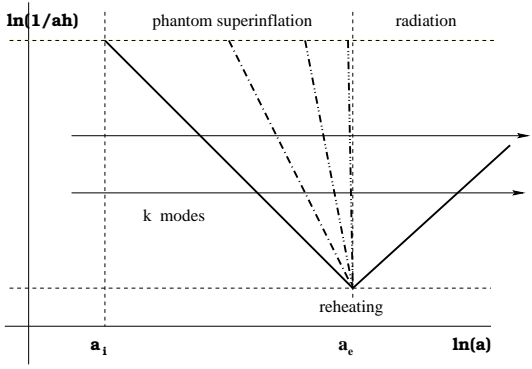}
\caption{ The sketch of evolution of $\ln{(1/ah)}$ with respect to
the scale factor $\ln{a}$ during the phantom superinflation for
different $\epsilon$, in which in the phantom superinflation
region the solid line denotes the usual inflation with
$\epsilon\simeq 0$, and the dot, two dots and three dots -dashed
lines denote the superinflation with $\epsilon = -1, -5, -\infty$,
respectively. The perturbation modes leave the Hubble horizon
during the phantom superinflation and then reenter the horizon
during the radiation/matter domination at late time. }
\end{center}
\end{figure}


We will discuss the primordial tensor perturbations from the
phantom superinflation in the following. In the momentum space,
the equation of motion of gauge invariant variable $u_k$, which is
related to the tensor perturbation by $u_k \equiv a h_k$, is \be
u_k^{\prime\prime} +(k^2-{a^{\prime\prime}\over a}) u_k = 0
,\label{uk}\ee where $a(\eta)$ is given by (\ref{aeta}).
The general solutions of this equation are the Hankel functions.
In the regime $k\eta \rightarrow \infty $, all interesting modes
are very deep in the horizon of phantom phase, thus (\ref{uk}) can
be reduced to the equation of a simple harmonic oscillator $u_k
\sim {e^{-ik\eta} \over (2k)^{1/2}}$, which in some sense suggests
that the initial condition can be taken as usual Minkowski vacuum.
In the superhorizon scale, in which the modes become unstable and
grow, the expansion of Hankel functions to the leading term of $k$
gives \be u_k\simeq
{1\over \sqrt{2k}}(-k\eta)^{{1\over 2}-\nu} ,\ee where the phase
factor and the constant with order one have been neglected, and
\be \nu^2\equiv {1\over 4}({\epsilon-3\over \epsilon-1})^2 .\ee In
the phantom phase in which $\epsilon<0 $, we have $1/2<\nu <3/2$.

The amplitude of gravitational wave after the end of phantom
superinflation is given by, accounting for both polarizations, \be
{\cal P}_t^{1/ 2}(k,\eta_e)= {k^{3/2}\over \pi}
|{u_k(-k\eta_e)\over a_e}|\simeq {(-\eta_e)^{1/2-\nu}\over a_e}
k^{3/2-\nu}.\ee We obtain, from (\ref{htau}) and (\ref{caln}), \ba
{\cal P}_t^{1/ 2}(k,\eta_e) & =& {1\over \pi a_e}({1\over
(1-\epsilon) a_e h_e})^{1/2-\nu} k^{3/2-\nu}\nonumber\\ &=&
{h_e\over \pi}{1\over (1-\epsilon)^{1/2-\nu}}
({k_0\over k_e})^{3/2-\nu}({k\over k_0})^{3/2-\nu}\nonumber\\
&=&{h_e \over \pi} {e^{-{\cal N}(3/2-\nu)}\over
(1-\epsilon)^{1/2-\nu}}({k\over k_0})^{3/2-\nu} . \label{pt}\ea
We can see that for fixed $\cal N$ and $h_e$, when $\nu \simeq
3/2$, which corresponds to $\epsilon \simeq 0$, ${\cal P}_t^{1/
2}(k,\eta_e)\simeq h_e/\pi $ is scale invariant, which is the
usual result of inflation models, while when $\nu <3/2$, which
corresponds to $\epsilon <0$, the spectrum of ${\cal P}_t^{1/
2}(k,\eta_e)$ is blue tilted and has the exponentially suppressed
amplitude $\sim h_e e^{-{\cal N}(3/2 -\nu)}/\pi $ at the largest
scale $k\simeq k_0$. The limit of $\nu\rightarrow 1/2$ is the
solution of Ref. \cite{PZ}, in which the scale factor is nearly
unchanged and the Hubble parameter $h$ experiences an
instantaneous ``jump", which results in that the spectrum is
strong blue tilt \be {\cal P}_t^{1/ 2}(k,\eta_e)\simeq h_e
e^{-{\cal N}}(k/k_0)/\pi .\ee

To convert from the primordial spectrum to the present spectrum,
we need to know the transfer function $T(k)$, \be {\cal P}_t^{1/
2}(k,\eta_0)=T(k) {\cal P}_t^{1/ 2}(k,\eta_e). \ee In general,
${\cal P}(k,\tau)$ is roughly time independent outside the horizon
and decay as $a^{-1}$ as long as the corresponding mode reenters
the horizon. Therefore, based on the fact that $h\sim a^{-2}$
during the radiation domination and $h\sim a^{-3/2}$ during the
matter domination, the numerical fitting gave \cite{TLW, T} \be
T(k) \simeq ({k_0\over k})^2 (1+{4\over 3}{k\over k_{eq}}+{5\over
2}({k\over k_{eq}})^2)^{1/2}, \label{tk}\ee where
$k_{eq}=a_{eq}h_{eq}$ denotes the modes entering the horizon at
the time of matter-radiation equality. Eq. (\ref{tk}) actually
only applies for $k>k_0$ and below this $T(k)=1$. We have, from
(\ref{asim}), \be {k_{eq}\over k_0}={a_{eq}h_{eq}\over a_0
h_0}=({a_0\over a_{eq}})^{\epsilon-1} . \label{keq}\ee For the
matter domination, $\epsilon= 3/2$. Thus $k_{eq}/k_0
=\sqrt{1+z_{eq}} $.

\begin{figure}[t]
\begin{center}
\includegraphics[width=8cm]{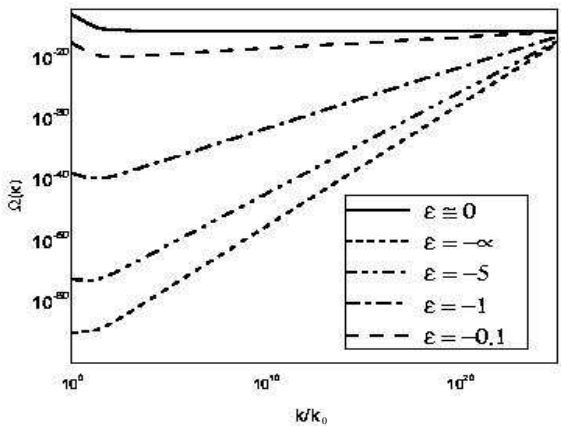}
\caption{The figure of $\Omega(k)$ from the phantom superinflation
with different $\epsilon$, where the horizon axis is $k/k_0$
taken from $1$ to $10^{25}$. We take $k_0=10^{-18}$ and
$h_e^2=10^{-12}$. Thus $k_{max}\sim 10^{8}$ and $k_{max}/k_0\sim
10^{26}$.  }
\end{center}
\end{figure}

To characterize the spectrum of the stochastic gravitational wave
signal, an useful quantity $\Omega_{gw}(k)$ is introduced, see
\cite{Th, A1, B}, which is the gravitational wave energy per unit
logarithmic wave number in the units of the critical density \be
\Omega_{gw}(k)= {k\over \rho_{cr} }{d\rho_{gw}\over d k} , \ee
where $\rho_{cr}= 3h_0^2/8\pi$ is the critical energy density of
observable universe and $\rho_{gw}$ is the energy density of
gravitational wave. The value of $h_0$ is generally written as
$h_0= {\tilde h}_0 \times 100$ km/sec/Mpc, where ${\tilde h}_0$
parameterizes the existing experimental uncertainty. However, for
simplicity we will neglect ${\tilde h}_0$ in the following. In
general \cite{T}, \ba \Omega_{gw}(k) &\simeq &
{1\over 6} ({k\over k_0})^2 {\cal P}_t(k,\eta_0)\nonumber\\
&\simeq & {1\over 6}({k_0\over k})^2(1+{4\over 3}{k\over
k_{eq}}+{5\over 2}({k\over k_{eq}})^2){\cal P}_t(k,\eta_e) ,\ea
where (\ref{tk}) has been used. Fig.2 shows the results of
$\Omega_{gw}(k)$ for the phantom superinflation with different
$\epsilon$. The spectrum is cut off at the wave number
$k_{max}=k_e$. For the instantaneous reheating in which the
phantom superinflation is immediately followed by a phase of
radiation domination, $k_{max}=a_eh_e\simeq a_0 T_0 h_e^{1/2}$,
where $T_0$ is the temperature of present CMB. The overall
amplitude $\sim h_e^2$ of $\Omega_{gw}(k)$ should be given by the
CMB observation at the large angular scale. The present bound is
$\Omega_{gw}(k)<10^{-11}k_0/k$ at $10^{-18}$Hz $<k<10^{-16}$Hz
\cite{AK}. We can see that for $\epsilon\simeq 0$,
$\Omega_{gw}(k)$ decays with $k$ up to $k\simeq k_{eq}$, and after
$k\gg k_{eq}$ it enters a nearly long plateau \be \Omega_{gw}(k\gg
k_{eq},\epsilon\simeq 0)\simeq {5 h_e^2\over 12 \pi^2}({k_0\over
k_{eq}})^2, \ee which is the character of inflation model
\cite{T}, while for $\epsilon <0$ the case is distinctly
different, in which at $k\gg k_{eq}$, $\Omega_{gw}(k)$ will
increase with $k$ up to $k\simeq k_{max}$. The slope of increasing
is very dependent of $\epsilon$, and the extreme one is $\sim k^2$
corresponding to $\epsilon\rightarrow -\infty$ \cite{PZ}. However,
since at low frequency region, the amplitude of $\Omega_{gw}(k)$
has a strong suppression from the efolding number, see (\ref{pt}),
which is relevant with $\epsilon$, at the case that the reheating
temperature $T_r\simeq h_e^{1/2}$ is same we can
obtain at $k\gg k_{eq}$ \ba & & \Omega_{gw}(k_{max})\nonumber\\
&\simeq & (1-\epsilon)^{2\nu-1}{5 h_e^2\over 12 \pi^2}({k_0\over
k_e})^2({k_e\over k_{eq}})^2({k_0\over k_e})^{3-2\nu}({k_e\over
k_0})^{3-2\nu}\nonumber\\ & = &(1-\epsilon)^{2\nu-1}{5 h_e^2\over
12 \pi^2}({k_0\over k_{eq}})^2\nonumber\\ &\simeq &
\Omega_{gw}(k\gg k_{eq},\epsilon\simeq 0). \ea Thus though for
$\epsilon <0$ there is an interesting increasing of
$\Omega_{gw}(k)$ amplitude at high $k$, when the scale $h_e$ (or
reheating scale) is taken as same, the largest value obtained at
$k\simeq k_{max}= k_e$ is always around the plateau $
\Omega_{gw}(k\gg k_{eq},\epsilon\simeq 0)$ of the inflation. This
result is not dependent on the value of the reheating scale.

In summary, we study in detail the gravitational wave background
excited during the phantom superinflation. For $\epsilon\simeq
0_-$, the amplitude of spectrum is very close to that of
inflation, and can be expected to be detected in the future. But
for $\epsilon<< -1$, the amplitude is very low at the large scale
(low $k$) and hardly seen, however, in high frequency region the
amplitude increases $\sim k^{3-2\nu}=k^{2\epsilon/(\epsilon -1)}$
up to the plateau of gravitational wave background of the
inflation, thus there may be some significant and different
signals around $k_{max}$. These results may be interesting for the
present and planned experiments detecting gravitational wave.

\textbf{Acknowledgments} This work was partly completed in
Interdisciplinary Center of Theoretical Studies, Chinese Academy
of Sciences, and also in part by NNSFC under Grant No: 10405029.
90403032, and National Basic Research Program of China under Grant
No: 2003CB716300.


\begin{thebibliography}{99}

\bibitem{G} L.P. Grishchuk, Sov. Phys. JETP \textbf{40}, 409
(1975).

\bibitem{A} B. Allen, Phys. Rev. \textbf{D37}, 2078 (1988).

\bibitem{S} A. Starobinski, JETP Lett. \textbf{30}, 682 (1979).


\bibitem{RSV} V. Rubakov, M. Sazhin and A. Veryaskin, Phys. Lett.
\textbf{115B}, 189 (1982).

\bibitem{FP} R. Fabbri and M.D. Pollock, Phys. Lett.
\textbf{125B}, 445 (1983).

\bibitem{S1} A. Starobinski, Sov. Astron. Lett. \textbf{9}, 302
(1983).

\bibitem{AW} L. Abbott and M. Wise, Nucl. Phys. \textbf{B244}, 541
(1984).

\bibitem{AH} L. Abbott and D. Harari, Nucl. Phys. \textbf{B264},
487 (1986).

\bibitem{GG} M. Gasperini and M. Giovannini, Phys. Rev.
\textbf{D47}, 1519 (1993).

\bibitem{BGGV} R. Brustein, G. Gasperini, G. Giovannini and G.
Veneziano, Phys. Lett. \textbf{B361}, 45 (1995).

\bibitem{BMU} A. Buonanno, M. Maggiore and C. Ungarelli, Phys.
Rev. \textbf{D55}, 3330 (1997).

\bibitem{BST} L.A. Boyle, P.J. Steinhardt and N. Turok, Phys.
Rev.  \textbf{D69}, 127302 (2004).

\bibitem{BEPS} B. Boisseau, G. Esposito-Farese, D. Polarski, A.A. Starobinsky, Phys. Rev. Lett. 85, 2236 (2000).

\bibitem{PZ} Y.S. Piao and E Zhou, Phys. Rev. \textbf{D68}, 083515
(2003).

\bibitem{PZ1} Y.S. Piao and Y.Z. Zhang, Phys. Rev. \textbf{D70},
063513 (2004).

\bibitem{GJ} P.F. Gonzalez-Diaz and J.A. Jimenez-Madrid, Phys. Lett.
\textbf{B596}, 16 (2004).

\bibitem{ABV} A. Anisimov, E. Babichev and A. Vikman,
astro-ph/0504560.

\bibitem{BFM} M. Baldi, F. Finelli and S. Matarrese,
astro-ph/0505552.

\bibitem{NO} S. Nojiri and S.D. Odintsov, hep-th/0506212;
hep-th/0507182.

\bibitem{PZ2} Y.S. Piao and Y.Z. Zhang, Phys. Rev. \textbf{D70}, 043516
(2004); Y.S. Piao, Phys. Lett. \textbf{B606} (2005) 245.

\bibitem{KST} J. Khoury, P.J. Steinhardt and N. Turok,
astro-ph/0302012.

\bibitem{TLW} M.S. Turner, J.E. Lidsey and M. White, Phys. Rev.
\textbf{D48}, 4613 (1993).

\bibitem{T} M.S. Turner, Phys. Rev. \textbf{D55}, 435 (1997).

\bibitem{Th} K.S. Thorne, in ``300 Years of Gravitation", edited
by S. Hawking and W. Israel.

\bibitem{A1} B. Allen, gr-qc/9604033.

\bibitem{B} A. Buonanno, gr-qc/0303085.

\bibitem{AK} B. Allen and S. Koranda, Phys. Rev. \textbf{D50},
3713 (1994).

\end{thebibliography}
\end{document}